\documentclass[showpacs,superbib,10pt,preprint,final,twocolumn,aps]{revtex4}
\usepackage{graphicx}

\newcommand{\bra}[1]{\langle #1 |}
\newcommand{\ket}[1]{| #1 \rangle}
\def\br{{\bf r}}
\def\ri{{\rm i}}
\def\re{{\rm e}}
\def\bea{\begin{eqnarray}}
\def\eea{\end{eqnarray}}
\def\ben{\begin{equation}}
\def\een{\end{equation}}

\begin{document}

\title{On the density-potential mapping in time-dependent density-functional theory}
\author{N. T. Maitra}
\affiliation{Department of Physics and Astronomy, Hunter College and the Graduate Center of the City University of New York, 695 Park Ave, New York, NY 10065, USA}
\author{T. N. Todorov}
\affiliation{School of Mathematics and Physics, Queen's University Belfast, Belfast BT7 1NN, UK}
\author{C. Woodward}
\affiliation{Department of Mathematics, Hill Center, Rutgers, the State University of New Jersey, 110 Frelinghuysen Road, Piscataway NJ 08854, USA}
\author{K. Burke}
\affiliation{Department of Chemistry, 1102 Natural Sciences 2, University of California Irvine, CA 92697, USA}

\pacs{PACS: 31.15.ee, 31.15.ec, 71.15.Qe}

\begin{abstract}
The key questions of uniqueness and existence in time-dependent density-functional
theory are usually formulated only for potentials and densities that are analytic 
in time. 
Simple examples, standard in quantum mechanics, lead however to non-analyticities.  We 
reformulate these questions in terms of a non-linear Schr\"odinger equation 
with a potential that depends non-locally on the wavefunction.
\end{abstract}

\maketitle

\section{Introduction}
The reduced one-particle probability density (henceforth density {\it tout court}) 
in a gas of identical interacting particles in an
external confining potential carries a wealth of information.  For example, 
the linear response of the density to a driving field is related
to density-density correlations and the excitation spectrum of the system. 
By continuity, the density evolution sets the total current through an
arbitrary open surface across a finite system driven out of
equilibrium, giving access to transport properties, or across a closed
surface, giving access to transfer processes out of a chosen volume. 
The density determines the generalized mean forces on macroscopic
degrees of freedom, with which the gas interacts, and hence the
coupled dynamics of the two subsystems.

In fact, the theorems of time-dependent (TD) density-functional theory
(DFT) prove that, given an initial many-body state, the
TD density contains in principle the expectations of {\it all} observables of
the interacting system, evolving under a TD scalar potential.
Following (the much older) ground-state DFT~\cite{HK64,KS65}, 
TDDFT calculations operate
by constructing a fictitious system of non-interacting electrons,
evolving under another TD scalar potential, such that the fictitious system reproduces
the electron density of the interacting system.  TDDFT thus provides access to the
properties of the interacting system via the much more tractable
corresponding non-interacting problem.  In recent years, TDDFT has
become a central tool for a range of problems involving
departures from the electronic ground state (GS).  Most applications are in the
linear-response regime, calculating the electronic excitation spectra and 
response of atoms, molecules, solids, even biomolecules, exposed to 
external fields~\cite{EFB07}.  Real-time electron dynamics in strong fields have 
also been studied, including electron transfer processes and transport, 
ionization, high-harmonic generation, and coupled non-adiabatic electron-nuclear
dynamics~\cite{TDDFTbook}.

TDDFT is based on two distinct results.  The first is that, under a
given particle interaction $\hat{W}$ and for a given initial
many-body state $\ket{\Psi(0)}$, there is a 1:1
correspondence between the ensuing evolution of the density and the
external potential acting on the system.  
This is the Runge-Gross
theorem (RG)~\cite{RG}. 
It implies that the potential and all other properties of the system
are {\em functionals} of the density and initial state, and practical calculations hope
to approximate some of these functionals accurately.

The second result is that for every density 
$n(\br,t)$, evolving from a given initial state $\ket{\Psi(0)}$ under a
given particle interaction $\hat{W}$ and external potential 
$v(\br,t)$, there exists an external potential $v'(\br,t)$ that returns the 
same density, $n(\br,t)$, under another particle interaction 
$\hat{W}'$ (the key case of interest being ${\hat W}' = 0$), starting
from, in general, another initial state $\ket{\Psi'(0)}$ (so long
as the two initial states share the same $n(\br,0)$ and $\dot{n}(\br,0)$). 
This result is van Leeuwen's $v$-representability
theorem (vL)~\cite{Robert1}, later extended to TD current DFT~\cite{Giovanni}.
It means that, by contrast with GS DFT, there is
no doubt about the existence of a KS system, i.e. a set of non-interacting
electrons whose TD density matches that of the interacting
system.  The $v$-representability difficulty is lessened in the TD
case because the TD Schr{\" o}dinger equation (TDSE) is first-order in
time, and the dependence on the initial wavefunction takes care of a large 
part of the difficulties associated with $v$-representability.

The first result has been
proven for potentials that are analytic in time about the initial time 
(henceforth denoted $t$-analytic), i.e. such that $v(\br,t)$ is equal to 
its Taylor series expansion in $t$ about the initial time, for a finite time interval.
The second assumes $t$-analytic
potentials {\it and} densities.  There are also two extensions of RG  in the 
linear-response regime that go beyond these analyticity requirements.
In the first~\cite{NS}, the short-time
density response to ``small'' but arbitrary potentials has been shown
to be unique under two assumptions: that the system starts from a
stationary state (not necessarily the GS) of the
initial Hamiltonian and that the corresponding linear density-response
function is $t$-analytic.  In the second~\cite{Robert2}, uniqueness
of the linear density response, starting from the electronic GS, was proven
for any Laplace-transformable (in time) potential.  As most 
physical potentials have finite Laplace transforms, this represents a
significant widening of the class of potentials for which a 1:1
mapping can be established  in the 
linear-response regime, from an initial GS; this includes e.g. potentials turning on as $\re^{-C/t^n}$ 
with $C>0$, $n>0$, or $t^p$ with $p$ positive non-integer.  It covers most cases of 
physical interest, under these conditions. 

The TDSE is a coupled partial differential
equation, which inextricably links analyticity properties in space and
time.  Thus potentials and initial wavefunctions that are (or could be) standard
examples in textbooks, in which the potential or the initial wavefunction
has some non-analytic properties
in space, yield densities that are not analytic in time.  We demonstrate
this with several examples in the last part of this paper.  This implies
that the assumptions underlying vL are
more restrictive than they may first appear.  

In this paper we build on the foundational work in Refs.~\cite{RG,Robert1} on
 the problems of uniqueness
and existence in TDDFT.   Our goal is to extend the fundamental framework laid there, by suggesting a simultaneous formulation of these two problems, which overcomes the requirement that 
the potential or density be analytic in time.  This formulation is
close to (but different from) several problems whose solutions have
been proven to exist mathematically.  We relate the
one-to-one mappings to the existence and uniqueness of solutions to a
particular type of non-linear TDSE (NLSE).  While there has been a
considerable amount of work on the NLSE~\cite{Cazenave},
the particular non-linear structure we have here has not been
investigated before, to the best of our knowledge.
We do not give a general solution to this problem, but discuss the features 
of the equation in relation to other NLSEs in the literature. 
As the existence of solutions of the latter do not require $t$-analyticity, 
this suggests that the reformulation will be a useful tool for exploring the
limitations (or extensions) of our knowledge of the domain of the
Runge-Gross functional.

In  Section II, we carefully introduce our reformulation.
We explore some consequences of the reformulation and
what is known about solutions to the NLSE that
we find.  Section III turns to the motivation behind 
this endeavor, with the aid of simple examples where the density 
evolution under $t$-analytic potentials is non-analytic in time.

\section{Alternative Formulation of the Density-Potential Mapping}
\label{sec:alternative}
It is straightforward to see that, for a given initial state
$\ket{\Psi(0)}$ for a system of $N$ electrons, a given {\it state}-evolution, 
$\ket{\Psi(t)}$, can be generated by at most one
TD scalar potential $v(\br,t)$, as follows from TDSE
\ben
\ri\hbar\,\ket{\dot{\Psi}(t)} = \hat{H}(t)\,\ket{\Psi(t)} = \left(\hat{T} + \hat{W} + \hat{v}(t)\right)\,\ket{\Psi(t)}\,
\label{eq:TDSE}
\een
as long as the wavefunction $\Psi({\bf r}_{1},...,{\bf r}_{N},t)$ is non-vanishing on a 
dense set of values of its arguments.  Here, $\hat{T}$, $\hat{W}$ and $\hat{v}(t)$ 
are the many-electron operators for the total kinetic energy, the
electron-electron interaction (with $\hat{W}=0$ corresponding to
non-interacting electrons) and $v(\br,t)$, respectively.  That no two
distinct $\hat{v}$'s can generate the same $\ket{\Psi(t)}$, up to a physically irrelevant purely time-dependent phase, can be
shown by assuming otherwise, inserting into TDSE, and finding
$\hat{v}_1 (t) - \hat{v}_2 (t)= c(t)$, a physically irrelevant purely time-dependent constant. 

What RG proves is the much stronger statement that not only the states must be different, 
but also their one-particle densities, 
\ben
n(\br,t) = \bra{\Psi(t)}\hat{n}(\br)\ket{\Psi(t)} \equiv n_{\Psi}(\br, t)\,,
\label{density}
\een
must differ, under 
the restriction that the potentials considered are analytic around $t=0$. Here $\hat{n}(\br)=\sum_{i=1}^{N}\,\delta(\br - \hat{\br}_{i})$ is the number-density operator
for the system of $N$ electrons with position operators $\{ \hat{\br}_{i} \}$.
This restriction means that potentials that switch on at $t\geq 0$
like $\re^{-C/t^n}$, $t^p$ with $p$ non-integer, or $t^n \ln(t)$ are
not covered by the proof.  (Note that the first example is infinitely
differentiable, with vanishing derivatives at $t=0$, while higher-order
derivatives of the second and third types diverge as $t\rightarrow
0^{+}$).  These potentials are Laplace-transformable in $t$, and in
the linear-response regime starting from the GS are covered by the vL extension 
of RG~\cite{Robert2}.

Here, we formulate (but do not prove) a 1:1 mapping defined via the existence and
uniqueness properties of the solution to a system of coupled equations.
Differentiating the continuity equation ($\dot{n}(\br,t) +\nabla\cdot {\bf j}(\br,t) = 0$, 
where ${\bf j}(\br,t)$ is the current density) once with respect to $t$ gives~\cite{Robert1}
\ben
\frac{1}{m}\,\nabla\cdot[ n_{\Psi}(\br,t)\nabla v(\br,t)]   =   \ddot{n}_{\Psi}(\br,t) +\nabla \cdot {\bf a}[\Psi(t)](\br,t) \,,
\label{SL0}
\een
where
\ben
{\bf a}[\Psi(t)](\br,t) = \frac{1}{\ri\hbar}\,\bra{\Psi(t)} [{\hat {\bf j}}(\br),{\hat T}+{\hat W}] \ket{\Psi(t)}
\label{a}
\een
with ${\hat {\bf j}}(\br)$ the current-density operator~\cite{qfootnote}.

We now choose a given {\it prescribed} density, which we denote by $n_{\rm aim}(\br,t)$, 
and a given initial many-electron state $\ket{\Psi(0)}$, such that 
$n_{\Psi}(\br,0)=n_{\rm aim}(\br, 0)$, $\dot{n}_{\Psi}(\br,0)=\dot{n}_{\rm aim}(\br,0)$. 
We then ask, does a potential $v(\br,t)$ exist, such that the density it generates via 
Eqs. (\ref{eq:TDSE}) and (\ref{density}), from the chosen initial state, is equal to the  
prescribed density, $n_{\Psi}(\br,t)=n_{\rm aim}(\br,t)$, and is this potential unique?

Noting the appearance of the time-evolving state $\ket{\Psi(t)}$ in the evaluation of 
${\bf a}[\Psi(t)](\br,t)$ in Eq.~(\ref{SL0}), we re-phrase the question by asking whether the 
{\it coupled} equations
\begin{eqnarray} 
&&\frac{1}{m}\,\nabla\cdot[ n_{\rm aim}(\br,t)\nabla v(\br,t)] =   \ddot{n}_{\rm aim}(\br,t) +\nabla \cdot {\bf a}[\Psi(t)](\br,t) \label{SL}  \\
&&\ri\hbar \,\ket{\dot{\Psi}(t)} = \hat{H}(t)\,\ket{\Psi(t)} \label{TDSE}
\end{eqnarray}
have a unique solution for $v(\br,t)$ and $\ket{\Psi(t)}$, given the inputs $\ket{\Psi(0)}$ 
and $n_{\rm aim}(\br,t)$.  
At a notional level, these coupled equations work as follows. 
Eq.~(\ref{SL}) determines the potential $v(\br,t)$, at every
$t$, as an instantaneous functional of the state $\ket{\Psi(t)}$; the
(fixed-time) equation is of Sturm-Liouville form, for which a unique locally square-integrable
solution exists for strictly positive densities, and locally
square-integrable right-hand side~\cite{Hormander}.
The restriction to nodeless densities is in practice a minor one as
most densities of $N$-electron systems do not have nodes; but it
certainly excludes a small subset of them, and excludes typical
TD and excited-state densities of one-electron systems.
The restriction to locally square-integrable right-hand sides (meaning
that the integral of the square of the right-hand side over a finite
region is finite) is generally satisfied for physical wavefunctions~\cite{L99fnote}.
A formal solution to Eq.~(\ref{SL}) is discussed in the appendix.
Eq.~(\ref{TDSE}) then determines the evolution of the state, under the potential just
obtained, through an infinitesimal time-step to the ``next'' time, and
so on.  Eqs.~(\ref{SL}) and (\ref{TDSE}) therefore {\it define} the
density-potential mapping: if a unique solution $\left(v(\br,t),
\ket{\Psi(t)}\right)$ to Eqs.~(\ref{SL}) and (\ref{TDSE}) exists for a
given $\left(\ket{\Psi(0)},n_{\rm aim}(\br,t)\right)$, then there is a
1:1 mapping between the density and potential.  
If, further, the density $n_{\rm aim}(\br,t)$ is $v$-representable, it directly 
follows that the unique solution $\left(v(\br,t),\ket{\Psi(t)}\right)$
reproduces this target density, $n_\Psi(\br,t) = n_{\rm aim}(\br,t)$. 

Non-$v$-representability can arise in two ways. First, if no solution exists to Eqs.~(\ref{SL})-(\ref{TDSE}) for a particular $n_{\rm aim}(\br,t)$, then $n_{\rm aim}(\br,t)$ is not $v$-representable. 
Second,  a solution to systems of equations of type~(\ref{SL})-(\ref{TDSE}) does not necessarily have $n_\Psi(\br,t) \neq n_{\rm aim}(\br,t)$~\cite{Rauch}. If there does exist a unique solution $\left(v(\br,t),\ket{\Psi(t)}\right)$ but one that
fails to return the chosen target density, $n_\Psi(\br,t) \neq n_{\rm aim}(\br,t)$, then this target density is not $v$-representable~\cite{note1}.



The coupled equations~(\ref{SL}) and ~(\ref{TDSE}) can be expressed as evolution equations as follows. By taking one further $t$-derivative of Eq.~(\ref{SL}),
we obtain the system of first-order, in time, partial differential equations 
\bea
\nonumber
\frac{1}{m}\,\nabla\cdot[n_{\rm aim}(\br,t)\nabla\dot{v}(\br,t)] &=& 
- \frac{1}{m}\,\nabla\cdot[\dot{n}_{\rm aim}(\br,t)\nabla v(\br,t)] \\
\nonumber
&+&  \dddot{n}_{\rm aim}(\br,t) + \nabla \cdot \dot{{\bf a}}[\Psi(t)](\br,t)  \\
\ri\hbar\,\ket{\dot{\Psi}(t)} &=& \hat{H}(t)\,\ket{\Psi(t)}
\label{eq:dvSLdt}
\eea
for the pair $\left(v(\br,t),\ket{\Psi(t)}\right)$.
The initial conditions are the initial state $\ket{\Psi(0)}$ and 
$v(\br,0)$, given by the solution of
\ben
\frac{1}{m}\,\nabla\cdot[n_{\rm aim}(\br,0)\nabla v(\br,0)] 
= \ddot{n}_{\rm aim}(\br,0) +\nabla \cdot {\bf a}[\Psi(0)](\br,0) \,.
\label{eq:dvSLdtic}
\een 
A system of first-order ordinary differential equations with
fixed initial conditions always has a unique solution (given
appropriate continuity conditions on the
right~\cite{picardlindeloef}, such as Lipschitz continuity).  But 
partial differential equations are more complex.  While
the initial-value problem for many
classes of partial differential equations is known to have a unique
solution, to our knowledge, there are no results yet for 
equations of the form~(\ref{eq:dvSLdt}).



Instead, we consider a reformulation of
the coupled equations (\ref{SL}) and (\ref{TDSE})
as a NLSE with non-local potential term, i.e. the potential at $\br$ 
depends on the wavefunction not just at or near $\br$ but further away too.
The solution for $v(\br,t)$  is
an instantaneous functional of 
$\ddot{n}_{\rm aim}(\br,t)$, $n_{\rm aim}(\br,t)$ and $\ket{\Psi(t)}$ as can be seen from Eq.~(\ref{SL}), 
and its dependence on $\ket{\Psi(t)}$ is quadratic.
Denoting this solution symbolically by
$v(\br,t)=v[\ddot{n}_{\rm aim}, n_{\rm aim}, \Psi](\br,t)$,
Eqs.~(\ref{SL}) and (\ref{TDSE}) now reduce to the effective NLSE \ben \ri\hbar\,\ket{\dot{\Psi}(t)} = \left({\hat T}+{\hat W}+\hat
{v}[\ddot{n}_{\rm aim}, n_{\rm aim}, \Psi](t)\right)\,\ket{\Psi(t)}\,.
\label{TDSEf}
\een
The problems of $v$-representability of the chosen density 
$n_{\rm aim}(\br,t)$ and of the uniqueness of this representability 
reduce to the problem of the existence and uniqueness of
the solution to Eq.~(\ref{TDSEf}) (together with the requirement that
the solution returns  $n_\Psi(\br,t) = n_{\rm aim}(\br,t)$, whose
negation implies that the chosen density $n_{\rm aim}(\br,t)$ is not
$v$-representable, as discussed earlier).  
Two specific issues that bear on this problem
are the nature of the non-linearity of this equation and the boundary behaviour 
of ${\hat v}[\ddot{n}_{\rm aim}, n_{\rm aim},\Psi](t)$.

NLSEs with non-local interaction were investigated by Ginibre
and Velo \cite{GinibreVelo}, who in particular proved under certain conditions
the existence and uniqueness of
short-time solutions.  However their non-linear term (a) does not
include Coulomb interactions, (b) deals only with the version of Eq.~(\ref{SL}) 
for the ordinary Laplacian, and (c) does not involve derivatives of the wavefunction. 
Much of the recent mathematical work
involves non-linearities depending on
the wavefunction and a single derivative, whereas in this case the
potential depends non-locally on four derivatives of $\Psi$. 
For point (b), the results of Ref.~\cite{GinibreVelo} could be extended 
to a general uniformly elliptic operator on the left (i.e. one which has 
eigenvalues bounded below by a positive constant, everywhere in space), 
but this is not what we have:  the density appearing on the left-hand side 
of Eq.~(\ref{SL}) decays at infinity, so the differential operator on the left 
is not uniformly elliptic (i.e. $n_{\rm aim}(\br,t)$ is not
gapped away from zero).  This makes it difficult to obtain estimates on
the behavior of $v(\br,t)$ at infinity needed to reproduce the Picard
iteration needed to show existence and uniqueness as in 
Ref.~\cite{GinibreVelo}.  Problem (b) is absent in the
simplified case of periodic boundary conditions, and smoothing over
Coulomb singularities deals with problem (a) (as would be relevant for solid-state calculations 
using pseudo-potentials with a smoothed-out interaction term).  For example, Burq,
Gerard, and Tzvetkov \cite{BGT} prove short-time existence and
uniqueness of the NLSE with local
potential term on compact manifolds.  However, even in this case it is not clear
whether the given NLSE has the hoped-for
solution, because of the loss of derivatives (point (c) above).  Ref.~\cite{KPV} proves well-posedness for a NLSE that loses two or more derivatives, however for Hamiltonians with local interactions only. 
Part of the motivation for this paper is the hope that the further analysis of this
problem in mathematical physics will stimulate future work.


\section{Examples of non t-analytic densities}
\label{sec:examples}
The practical motivation behind a formulation that goes beyond the
analyticity requirements of the original Runge-Gross and van
Leeuwen theorems is that it is not unusual, and certainly not
pathological, for temporal non-analyticity to arise.  If it were true 
that under any particle interaction, a $t$-analytic potential always generates a $t$-analytic
density, and vice-versa,
then the constructs of Refs.~\cite{RG} and~\cite{Robert1} 
would be sufficient to provide a {\it self-contained} 
(even if somewhat restrictive) framework for TDDFT.  However,
we now argue, with the aid of several examples, that in 
general a $t$-analytic potential need not produce a $t$-analytic
density.  Although the RG 1:1 mapping holds in such cases, the
$v$-representability proof of Ref.~\cite{Robert1} does not apply, 
opening up the question of the existence and uniqueness of a corresponding Kohn-Sham system.

\subsection{Spatially non-analytic initial states}

In the first set of examples, spatial non-analyticity in the initial
wavefunction leads to temporal non-analyticity in the density.

\subsubsection{``Bump'' initial states: smooth but non-analytic}

Consider a free particle in 1d with an initial wavefunction $\psi(x,0)$ that is 
constant in space within a region, e.g. 
\ben
\psi(x,0) = \left\{
\begin{array}{lll} f(x)\,\,\,{\rm for}\,\,\,x<a\\C\,\,\,{\rm for}\,\,\,a\leq x \leq b\\g(x)\,\,\,{\rm for}\,\,\,x>b\,.\end{array}\right.\label{phi}
\een
Here, $C$ is a constant, and $f(x)$ and $g(x)$ are assumed to be such that $\psi(x,0)$ 
is infinitely differentiable in $x$ everywhere. 
By repeated application of TDSE
\ben
\ri\hbar\,\dot{\psi}(x,t) = -\frac{\hbar^{2}}{2m}\,\psi''(x,t)\,,
\label{freeTDSE}
\een
we see that, for a point $x=x_{0}$ in $(a,b)$, all $t$-derivatives 
of $\psi(x_{0},t)$ vanish at $t=0$~\cite{note3}. Thus, the Taylor series in time for $\psi(x_{0},t)$ at fixed $x_0$ is 
\ben
\psi_{T}(x_{0},t) = \sum_{k=0}^{\infty}\,\frac {t^k}{k!}\, \partial_{t}^{k}\psi(x_0,t)\vert_{t=0}=C\,. 
\een
At the points $x_{0}$ considered, this series exists, and converges (with, furthermore, 
an infinite radius of convergence), but not to the function itself:
$\psi_{T}(x_{0},t) = C \neq \psi(x_{0},t)$.  The true wavefunction, $\psi(x_{0},t)$, evolves in
time, since the initial state is not a free-particle eigenstate. In particular, the true solution will have density from the outer regions, $x<a$ and $x>b$, dispersing into the inner region, $a<x_0<b$. 
Therefore, $\psi(x_0,t)$ is not analytic in time.  Similarly, using
\bea
\partial_{t}^{k}n(x_0,t)\vert_{t=0} &=& \sum_{l=0}^{k}\,\frac{k!}{(k-l)!l!}\,\times \nonumber \\
&\times& \partial_{t}^{k-l}\psi^{*}(x_0,t)\vert_{t=0}\partial_{t}^{l}\psi(x_0,t)\vert_{t=0}\,,
\eea
for the Taylor series for the density at $x_{0}$ we obtain $n_{T}(x_{0},t) =
|C|^{2}$; it exists and converges, but not to $n(x_{0},t)$: $n_{T}(x_{0},t) \neq n(x_{0},t)$. 
Thus, the density is not $t$-analytic, even though the potential ($v(x,t)\equiv 0$) is.

The above arguments may be straightforwardly extended to initial states that have compact 
support but are smooth: there in the region $(a,b)$, $\psi(x,0) = f(x)$ while outside
this region, $\psi(x,0) = 0$.  Such functions are sometimes referred to as ``bump functions''. The true solution $\psi(x,t)$  disperses, while the time Taylor series of $\psi(x,t)$ does not.
Although formally solving the TDSE, the Taylor series  is not a valid solution because 
it is not uniformly convergent in the non-constant region.
Similar observations and a detailed analysis for free propagation of smooth compactly supported initial states can be found in Ref.~\cite{HS72}.

The above example is easily extended to any initial ``spline'' state 
\ben
\psi(x,0) = \left\{
\begin{array}{lll} f(x)\,\,\,{\rm for}\,\,\,x<a\\\psi_{n}(x)\,\,\,{\rm for}\,\,\,a\leq x \leq b\\g(x)\,\,\,{\rm for}\,\,\,x>b\,,\end{array}\right.
\een
where $\psi_{n}(x)$ is the $n$th stationary state, with energy $E_{n}$, of some static confining potential $v(x)$, 
and $f(x)$ and $g(x)$ are arbitrary (but may, if desired, be assumed to be such as to ensure infinite smoothness 
everywhere). Starting from this initial state, we consider evolution under the potential $v(x)$.
By repeated application of TDSE 
\ben
\ri\hbar\,\dot{\psi}(x,t) = -\frac{\hbar^{2}}{2m}\,\psi''(x,t) + v(x)\,\psi(x,t)\,,
\een
for $x_{0} \in (a,b)$ we arrive at 
$\psi_{T}(x_{0},t) = \re^{-\ri E_{n}t/\hbar}\,\psi_{n}(x_{0})=\re^{-\ri E_{n}t/\hbar}\,\psi(x_{0},0)$ 
and $n_{T}(x_{0},t) = |\psi_{n}(x_{0})|^{2}=n(x_{0},0)$, with the 
conclusion, again, that neither the wavefunction nor the density is $t$-analytic.

\subsubsection{Initial states with a cusp} 
\begin{figure}[h]
\centering
\includegraphics[height=6cm,width=8cm]{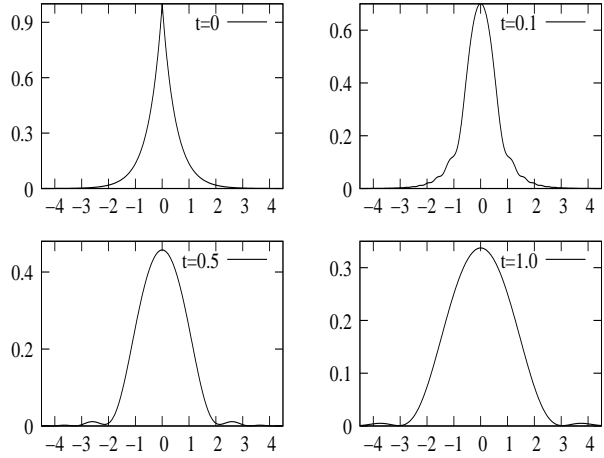}
\caption{The density of an initial exponential dispersing under free-particle evolution.}
\label{fig:dispersingGS}
\end{figure}

Consider evolving an initial state with a cusp in free space.
We can solve this explicitly in 1d, with:
\ben
\psi(x,0) = \sqrt{\kappa}\,\re^{-\kappa |x|}\,.
\label{cusp}
\een
Using the free-particle propagator
\ben
G(x,t; x',0) = \sqrt{\frac{m}{2\pi \ri \hbar t}}\, \re^{\ri (x-x')^2 m/2\hbar t}
\label{freeG}
\een
and the integral representation
\ben
\psi(x,t) = \int_{-\infty}^{+\infty}\,G(x,t; x',0) \psi(x',0)\,dx'\,,
\label{intrep}
\een
we obtain
\bea
\nonumber
\psi(x,t)&=& \frac{\sqrt{\kappa}}{2}\,\re^{\ri\hbar \kappa^2 t/2m}\left\{\re^{\kappa x}\left[1 - {\rm erf}\left(\frac{x +\ri\hbar \kappa t/m}{\sqrt{2\ri\hbar t /m}}\right)\right]\right. \\
&+& \left.\re^{-\kappa x}\left[1 - {\rm erf}\left(\frac{-x + \ri\hbar \kappa t/m}{\sqrt{2\ri\hbar t /m}}\right)\right]\right\}\,.
\label{eq:psidisapp}
\eea
Immediately from $t=0^+$, the cusp vanishes, and the initial exponential gradually disperses and develops 
oscillations in the tail (see Fig. \ref{fig:dispersingGS}, and note the diminishing scale on the vertical axis). 

From the Taylor series, away from $x=0$, we obtain simply
\ben 
\psi_T(x\neq0,t) = \sqrt{\kappa}\,\re^{\ri \hbar\kappa^2 t/2m}\,\re^{-\kappa \vert x\vert} \,,
\een
clearly not equal to the true wavefunction $\psi(x,t)$ (Eq.~(\ref{eq:psidisapp})). 
Noting that the true solution to TDSE is defined and continuous everywhere, we
define $\psi_{T}(0,t)$ such as to make $\psi_{T}(x,t)$ continuous. But then it obeys the TDSE
\ben
\ri\hbar\,\dot{\psi}_{T}(x,t) = - \frac{\hbar^{2}}{2m}\,\psi_{T}''(x,t) - \frac{\hbar^{2}\kappa}{m}\delta(x)\,\psi_{T}(x,t)\,,
\een
as opposed to the free-particle TDSE, Eq.~(\ref{freeTDSE}), obeyed by $\psi(x,t)$. Therefore the Taylor series gives a delta-function error in solving the TDSE.

The electron-nuclear Coulomb singularity in 3d gives rise to cusps in
stationary states and densities in real systems, i.e. initial states
are typically not smooth at the singularity.  For example, the
GS of the hydrogen atom, $\psi_H(\br) = \sqrt{\frac{1}{\pi a_0^3}}\, \re^{-r/a_0}$, where $a_0$ is the Bohr radius, has
analogous mathematical structure to the initial state of
Eq.~(\ref{cusp}). If made to evolve in a potential in which the Coulomb
singularity changes (e.g. a nucleus that vanishes, admittedly unphysical), the space-time coupling in the TDSE results in a
non-analytic evolving density, in a way that closely follows the above 1d
model. We need not however think of the initial state as a
GS of any potential: the initial state in Eq.~(\ref{cusp}) or
its 3d analog $\psi_H(\br)$ (or any of the examples given here) can be prepared in many different ways,
for example, using particular interferences of free-particle
eigenstates. The essential point is that we have a quantum-mechanically allowed initial state, which, when freely-evolved, gives a non-$t$-analytic density.

That non-$t$-analytic behavior is {\it not} a
consequence of a change or removal of a singularity in the potential,
is shown explicitly in the next example.

\subsubsection{Free evolution of the GS of a smooth bounded potential}
Consider a free particle in 1d evolving from
\ben
\psi(x,0) = \mathcal{N}\,(1-\re^{-C/x^{2}})\,,
\een
where $C>0$ is a (real) constant and $\mathcal{N} =1/\sqrt{2\sqrt{2\pi C}(\sqrt{2} -1)}$ is the 
normalisation constant. This initial wavefunction can be thought of as an example of the earlier 
bump states, with the interval $[a,b]$ shrunk to a point. It is the GS of  the potential $v(x) = \frac{\hbar^2}{m}\frac{C(3x^2 - 2C)}{x^6(e^{C/x^2} -1)}$, infinitely differentiable; see Fig. 2 for a plot. 
\begin{figure}[h]
\centering
\includegraphics[height=4cm,width=7cm]{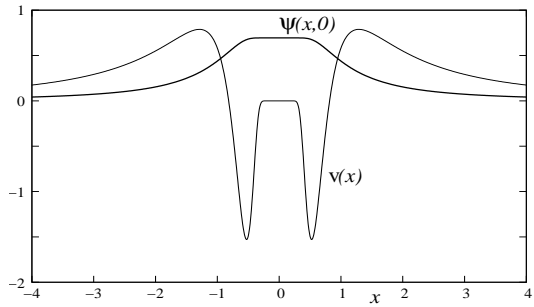}
\caption{The initial state of example A.3 (taking $C=1$), and the potential in which it is the ground state.}
\label{fig:expmxsq}
\end{figure}

Using Eqs.~(\ref{freeG}) and (\ref{intrep}), we find
\ben
\psi(0,t) = \mathcal{N}\,\left(1-\re^{-\sqrt{2mC/\ri\hbar t}}\right)
\een
and
\ben
n(0,t) = \mathcal{N}^2\,\left(1 - 2\,\re^{-\sqrt{mC/\hbar t}}\,\cos\sqrt{mC/\hbar t} + \re^{-2\sqrt{mC/\hbar t}}\right)\,.
\een
Neither is $t$-analytic.  The non-analyticity in the density takes the form of temporal evolution
that picks up infinitely slowly.

To summarize, the examples above illustrate how a non-$t$-analytic
density may arise from a (trivially) $t$-analytic potential, starting
from an initial state that is spatially non-analytic, even though it
could be smooth.  Indeed Ref.~\cite{HS72} points out that even for the simplest free-particle evolution, starting from an initially smooth state (our examples 1 and 3 are examples thereof), although the Taylor series solution is a formal solution of the TDSE, it is a valid solution only for a very restricted class of initial states, due to non-convergence. 
Even the time-evolution of the density in a hydrogen atom in an electric field can be shown to be non-$t$-analytic~\cite{YBunpub}.
The Runge-Gross theorem applies to all cases, as it is a
reformulation of the quantum dynamics of {\it any} initial state
allowed in quantum mechanics (satisfying appropriate decay at
infinity) under a $t$-analytic potential. But in cases where the ensuing evolution is non-$t$-analytic, the $v$-representability theorem does not apply.

\subsection{Spatially non-analytic potentials}

Finally, we consider potentials with the spatial non-analyticity of ``bump functions''.
We have one particle in 1d, starting from the $n$th stationary state $\psi_{n}(x)$, with
energy $E_{n}$, of some smooth static confining potential $v_{0}(x)$. Now evolve $\psi(x,t)$ 
(from $\psi(x,0)=\psi_{n}(x)$) under the potential 
\ben
v(x) = v_{0}(x) + v_{\rm ext}(x) \,,
\een
where
\ben
v_{{\rm ext}}(x) = \left\{
\begin{array}{lll} f(x)\,\,\,{\rm for}\,\,\,x<a\\C\,\,\,{\rm for}\,\,\,a\leq x \leq b\\g(x)\,\,\,{\rm for}\,\,\,x>b\,.\end{array}\right.\label{phieg2}
\een
Here, $C$ is a constant, and $f(x)$ and $g(x)$ may be chosen such as to give infinite differentiability in $x$ 
everywhere.  By applying TDSE repeatedly, for $x_{0} \in (a,b)$ we arrive at 
$\psi_{T}(x_{0},t) = \re^{-\ri (E_{n}+C)t/\hbar}\,\psi_{n}(x_{0})=\re^{-\ri (E_{n}+C)t/\hbar}\,\psi(x_{0},0)$ 
and $n_{T}(x_{0},t) = |\psi_{n}(x_{0})|^{2}=n(x_{0},0)$. Since both the wavefunction and 
the density must evolve, we conclude again that neither of them is $t$-analytic, even though the potential is.

\section{Summary and Outlook}
TDDFT is an exact reformulation of quantum mechanics in which the
density replaces the wavefunction as the basic variable. The
density-potential mapping is proven for any initial state evolving
under a $t$-analytic potential, such that the density decays
appropriately at infinity. The mapping to the KS system is proven for
$t$-analytic densities and potentials.  However, quite generally, the
densities arising from the TDSE are {\it not} $t$-analytic, as the
simple examples in Sec.~\ref{sec:examples} demonstrate. The initial
states and potentials considered are certainly within the realm of
traditional quantum mechanics, and nothing in the Runge-Gross theorem rules
them out from consideration for TDDFT.

The examples  suggest that in general a
$t$-analytic potential does not give a $t$-analytic density. However, then
it is no longer clear that a density, generated by a $t$-analytic
potential under one particle interaction, is represented by a
$t$-analytic potential under another.  This would throw into question the existence and uniqueness of the non-interacting KS system. Therefore, an extension of
the TDDFT framework, beyond $t$-analytic potentials {\em and}
densities, is necessary.  The present proposal is that Eqs.~(\ref{SL})
and (\ref{TDSE}) (or their two further formulations in Section
\ref{sec:alternative}) provide a route to one such extension.

The framing of the fundamental theorems of time-dependent density-functional
theories in terms of well-posedness of a type of NLSE first
appeared in Ref.~\cite{T09} where it arises naturally in Tokatly's
Lagrangian formulation of TD current-DFT, known as TD-deformation
functional theory. The traditional density-potential mapping question
is avoided in TD-deformation functional theory, where instead this
issue is hidden in the existence and uniqueness of a NLSE involving
the metric tensor defining the co-moving frame. Investigating the
relation between the particular NLSE of Ref.~\cite{T09} and the one in
the present paper is an interesting avenue for future work.

It is hoped that these brief considerations will motivate lively work
to take these preliminary developments further.

\section{Acknowledgements}
TNT is grateful to EPSRC for support through grant No. EP/C006739/1, and to Queen's University Belfast
for support through a Sabbatical. We also gratefully acknowledge financial support from the National Science Foundation grants CHE-0647913 (NTM) and DMS-060509 (CW), Department of Energy Grant DE-FG02-08ER46496 (KB) and a Research Corporation Cottrell Scholar Award (NTM). 

\section{Appendix}
In this appendix, we explore some interesting properties of the solution to Eq.~(\ref{SL}). Specifically, we discuss a reformulation that makes contact with the more standard theory of second-order elliptic operators, namely, operators that are uniformly elliptic at infinity. 

 We can
formally solve Eq.~(\ref{SL})
for $v(\br,t)$, with the aid of the transformation
\ben
w(\br,t) = v(\br,t)\sqrt{n_{\rm aim}(\br,t)}\,.
\label{w}
\een
Then Eq.~(\ref{SL}) becomes
\ben
\Lambda (t) w(\br,t) = - m\frac{s(\br,t)}{\sqrt{n_{\rm aim}(\br,t)}} \equiv u(\br,t)\,,
\label{Lambda}
\een
where
\bea
&&\Lambda (t)= -\nabla^{2} + \frac{1}{2}\,\frac{\nabla^{2}n_{\rm aim}(\br,t)}{n_{\rm aim}(\br,t)}
-\frac{1}{4}\,\left(\frac{\nabla n_{\rm aim}(\br,t)}{n_{\rm aim}(\br,t)}\right)^{2}\label{Lambda1}\\
&&s({\bf r},t) = \ddot{n}_{\rm aim}(\br,t) +\nabla \cdot {\bf a}[\Psi(t)](\br,t) \label{s}\,.
\eea
We must now solve this equation for $w(\br,t)$. In the following, $t$ may be treated as a parameter.  The operator $\Lambda(t)$ has the structure
of an ordinary one-particle Hamiltonian.  $\phi_{0}(\br,t) = \sqrt{n_{\rm aim}(\br,t)}$ 
is an eigenfunction of $\Lambda (t)$ with eigenvalue $0$.  Since $\int\,n_{\rm aim}(\br,t)\,d\br = N$,
$\phi_{0}(\br,t)$ is normaliseable (and thus obeys the boundary condition $\phi_{0}(\br,t)\rightarrow 0$
as $|\br|\rightarrow \infty$).  Therefore, $\phi_{0}(\br,t)$ is a bound state of $\Lambda(t)$.
Since the eigenvalue of $\phi_{0}(\br,t)$ is zero, for a solution to Eq.~(\ref{Lambda}) for $w(\br,t)$ to exist, 
it is necessary that the right-hand side be orthogonal to $\phi_{0}$, 
\ben
\int\,\phi_{0}(\br,t)\frac{s(\br,t)}{\sqrt{n_{\rm aim}(\br,t)}}\,d\br = \int\,s(\br,t)\,d\br = 0\,.
\label{condition}
\een
Via Eq.~(\ref{s}), this requires $\int_{S}\,{\bf a}[\Psi(t)](\br,t)\cdot d{\bf S}$ to vanish as the surface 
$S$ expands to infinity, which is a condition on the state $\ket{\Psi(t)}$ in the solution of 
Eqs.~(\ref{SL}) (or (\ref{Lambda})) and~(\ref{TDSE}).  Since the eigenvalue of $\phi_{0}(\br,t)$ is zero,
furthermore, Eq. (\ref{Lambda}) defines $w(\br,t)$ only up to an arbitrary additive amount of $\phi_{0}(\br,t)$.  
We choose this amount to be zero.  Denoting 
the eigenfunctions and eigenvalues of $\Lambda(t)$ by $\{\phi_{\nu}(\br,t)\}$ and
$\{\lambda_{\nu}(t)\}$, respectively, we can then uniquely solve Eq.~(\ref{Lambda}) for $w(\br,t)$:
\ben
w(\br,t) = \int\,{\cal G}(\br,{\bf r'},t) u(\br',t)\,d\br' \,,
\label{wsln}
\een
where
\ben
{\cal G}(\br,\br',t) = \sum_{\nu\neq 0}\,\phi_{\nu}(\br,t)\lambda^{-1}_{\nu}(t)\phi_{\nu}^{*}(\br',t)\,. \label{Lambdatildeinv}
\een
The potential $v(\br,t)$ is then obtained from Eq.~(\ref{w}).  The 
given solution for $w(\br,t)$ by construction is orthogonal to
$\phi_{0}(\br,t)$, and corresponds to a specific choice of the
additive spatial constant in $v(\br,t)$~\cite{note2}.  The boundary behaviour at infinity 
of $w(\br,t)$ in Eq.~(\ref{wsln}), and hence of $v(\br,t)$ in Eq.~(\ref{w}), is determined 
by ${\cal G}(t)$ and by the quantity $u(\br,t)$ in Eq.~(\ref{Lambda}).

No difficulty with this solution for $w(\br,t)$ arises, if the target density $n_{\rm aim}(\br,t)$ is nodeless.
Then $\phi_{0}(\br,t)$ too is nodeless, and since it is a bound state of $\Lambda(t)$, it must be the GS.
Then all other eigenvalues of $\Lambda(t)$ are positive, ${\cal G}(t)$ in Eq. (\ref{Lambdatildeinv}) exists, 
and the potential $v(\br,t)$ in Eq. (\ref{w}) is defined everywhere. 

The situation is more complicated if $n_{\rm aim}(\br,t)$ has nodes.  Then, first, $\Lambda(t)$ may have 
bound states degenerate with $\phi_{0}(\br,t)$ and the requirement of Eq. (\ref{condition}) must be extended
to all such states, which constitutes a stronger condition on $s(\br,t)$.  (That granted, the formal solution for $w(\br,t)$ 
is then still given by Eq. (\ref{wsln}), but with all bound states of $\Lambda(t)$, degenerate with $\phi_{0}(\br,t)$,
excluded from the definition of ${\cal G}(t)$ in Eq. (\ref{Lambdatildeinv}).)  Second, both $u(\br,t)$ in 
Eq. (\ref{Lambda}) and $v(\br,t)$ in Eq. (\ref{w}) may diverge at the nodes of $n_{\rm aim}(\br,t)$. 
One would then have to consider the nature of the singularity, and its effect on the propagating state vector.


\begin{references}
\bibitem{qfootnote}
In terms of density matrices, this can be evaluated as
\bea
{\bf a}[\Psi(t)](\br,t) &=&\frac{\hbar^{2}}{4m^{2}}\, (\nabla - \nabla')(\nabla^2 - \nabla'^2)\rho_1(\br',\br,t)\vert_{\br' =\br} \nonumber \\
&-&\frac{1}{m}\,\int P(\br,\br')\nabla  W(|\br -\br'|)\,d\br'
\eea
where $\rho_1$ is the one-body density matrix, $P$ is the pair-density (diagonal of the two-body density matrix) and $W(|\br -\br'|)$
is the electron-electron interaction potential.
\bibitem{note1} An alternative formulation is to replace the left-hand side of 
Eq.~(\ref{SL}) by $(1/m)\,\nabla\cdot[ n_{\Psi}(\br,t)\nabla v(\br,t)]$.
Subtracting Eqs. (\ref{SL0}) and (\ref{SL}) then gives $ \Delta \ddot{n}(\br,t) = 0$, 
where $\Delta n(\br,t)=n_{\Psi}(\br,t)- n_{\rm aim}(\br,t)$. 
The given initial conditions $\Delta n(\br,0) = 0$, $\Delta \dot{n}(\br,0) = 0$ then 
guarantee $\Delta n(\br,t)=0$ for all  $t \geq 0$. Existence and uniqueness of 
the solution, for $\left(v(\br,t),\ket{\Psi(t)}\right)$, to this modified formulation
is then a necessary and sufficient condition for unique $v$-representability of the chosen
target density $n_{\rm aim}(\br,t)$.  However, this formulation makes $v(\br,t)$, as defined by 
Eq.~(\ref{SL}), a more complicated functional of $\ket{\Psi(t)}$.  The distinction between these 
two formulations is left for future work.
\bibitem{note3} Here, we have assumed, further, that $t$- and $x$-differentiation may be interchanged, and that this interchangeability 
survives as $t\rightarrow 0^{+}$.
\bibitem{note2} Since Eq.~(\ref{Lambda}) is of second order, the complementary function in the general solution for $w(\br,t)$ 
contains a second term, in addition to $\phi_0(\br,t)$. It too has been eliminated in the solution in Eq.~(\ref{wsln}).
\end{references}
\end{document}